\documentclass[aps,reprint,prb,amsmath,amssymb]{revtex4-1}

\usepackage{graphicx}
\usepackage{dcolumn}
\usepackage{bm}
\usepackage{subfig}
\usepackage[utf8]{inputenc}
\usepackage{amsmath,amssymb}
\usepackage{epstopdf}
\usepackage{textcomp}
\usepackage{hyperref}
\usepackage{listings}
\usepackage{upgreek}

\newcommand{\be}{\begin{equation}}
\newcommand{\ee}{\end{equation}}
\newcommand{\bal}{\begin{align}}
\newcommand{\eal}{\end{align}}
\newcommand{\bear}{\begin{eqnarray}}
\newcommand{\eear}{\end{eqnarray}}
\newcommand{\nn}{\nonumber}

\begin{document}

\title{Comment on ``Adiabatic Expansion of Electron Gas in a Magnetic Nozzle''\\
by Kazunori~Takahashi, Christine~Charles, Rod~Roswell, and Akira~Ando,\\
Phys. Rev. Lett. \textbf{120}, 045001 (2018)}

\author{Todor~M.~Mishonov}
\email[E-mail: ]{mishonov@bgphysics.eu}
\affiliation{Institute of Solid State Physics, Bulgarian Academy of Sciences,
72 Tzarigradsko Chaussee Blvd., BG-1784 Sofia, Bulgaria}

\author{Iglika~M.~Dimitrova}
\affiliation{Faculty of Chemical Technologies, University of Chemical Technology and Metallurgy,
8 Kliment Ohridski Blvd., BG-1756 Sofia, Bulgaria}

\author{Albert~M.~Varonov}
\email[E-mail: ]{varonov@issp.bas.bg}
\affiliation{Georgi Nadjakov Institute of Solid State Physics, Bulgarian Academy of Sciences,\\
72 Tzarigradsko Chaussee Blvd., BG-1784 Sofia, Bulgaria}

\date{4 August 2020, 16:48}

\begin{abstract}
Heat capacities at fixed volume and pressure 
as a function of the degree of ionization 
are graphically depicted as functions of reciprocal temperature 
and ionization degree. 
The polytropic index is calculated as a function of the same variables;
as it is not constant a partially ionized plasma can be only approximately
polytropic fluid.
In parallel, it is launched the idea that Alfv\'en waves can be used to heat the plasma in a propulsion jet with magnetic nozzle.
\end{abstract}

\maketitle

In this experiment with argon plasma in magnetic nozzle the interpretation of the experiment has 
revealed\cite{Takahashi:18} a polytropic index $\gamma$ definitely smaller than monoatomic value
$\gamma_a=5/3$.
The authors mention that in many laboratory experiments and astrophysical plasmas it is often
observed nearly isothermal expansion with $\gamma\sim 1.0 - 1.2$.
The authors discuss limit and breakdown of adiabaticity and extension of classical thermodynamics for modern physical systems.

The purpose of the of the present comment is to suggest a conventional interpretation of 
this deviation of $\gamma/\gamma_a$ from one.
We suppose that ionization-recombination processes are the main reason of this deviation and will present the analytical result for the polytropic index
\be
\gamma\equiv\frac{\mathcal{C}_p}{\mathcal{C}_v}=\frac{v_s^2}{v_{_T}^2},\qquad
v_s^2=\left(\frac{\partial p}{\partial \rho}\right)_{\!\!\!s},\qquad
v_{_T}^2=\left(\frac{\partial p}{\partial \rho}\right)_{\!\!T}
\ee
of the important case of hydrogen plasma.
Here $T$ is the temperature for which we use energy units, $\rho$ is the mass density,
$\mathcal{C}_p\equiv\left(\partial w/\partial T\right)_p$ is the heat capacity at constant pressure $p$ per unit mass, 
$\mathcal{C}_v\equiv\left(\partial \varepsilon/\partial T\right)_\rho$
is the heat capacity per unit mass at fixed volume, $\varepsilon$ is the internal energy per unit mass and $w=\varepsilon+p/\rho$
is the enthalpy per unit mass.

\begin{figure}[h]
\includegraphics[scale=0.32]{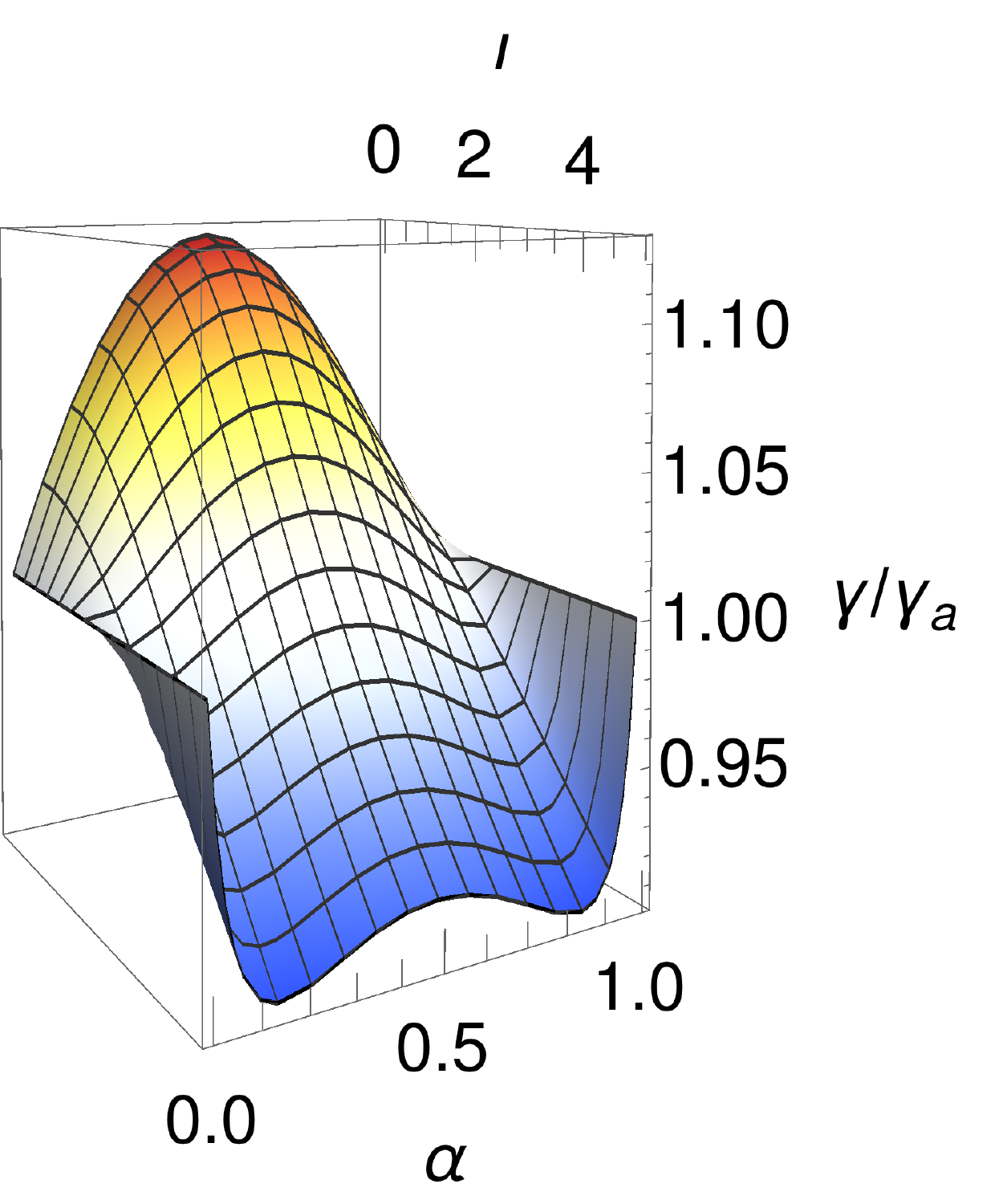}
\caption{
Analytical result for relative polytropic index $\gamma/\gamma_a$
as a function of  $\alpha$ and $\iota$.
The deviations from 1 are of order of the measured in the commented paper.\cite{Takahashi:18}
}
\label{Fig:gamma}
\end{figure}

\begin{figure}[h]
\includegraphics[scale=0.3]{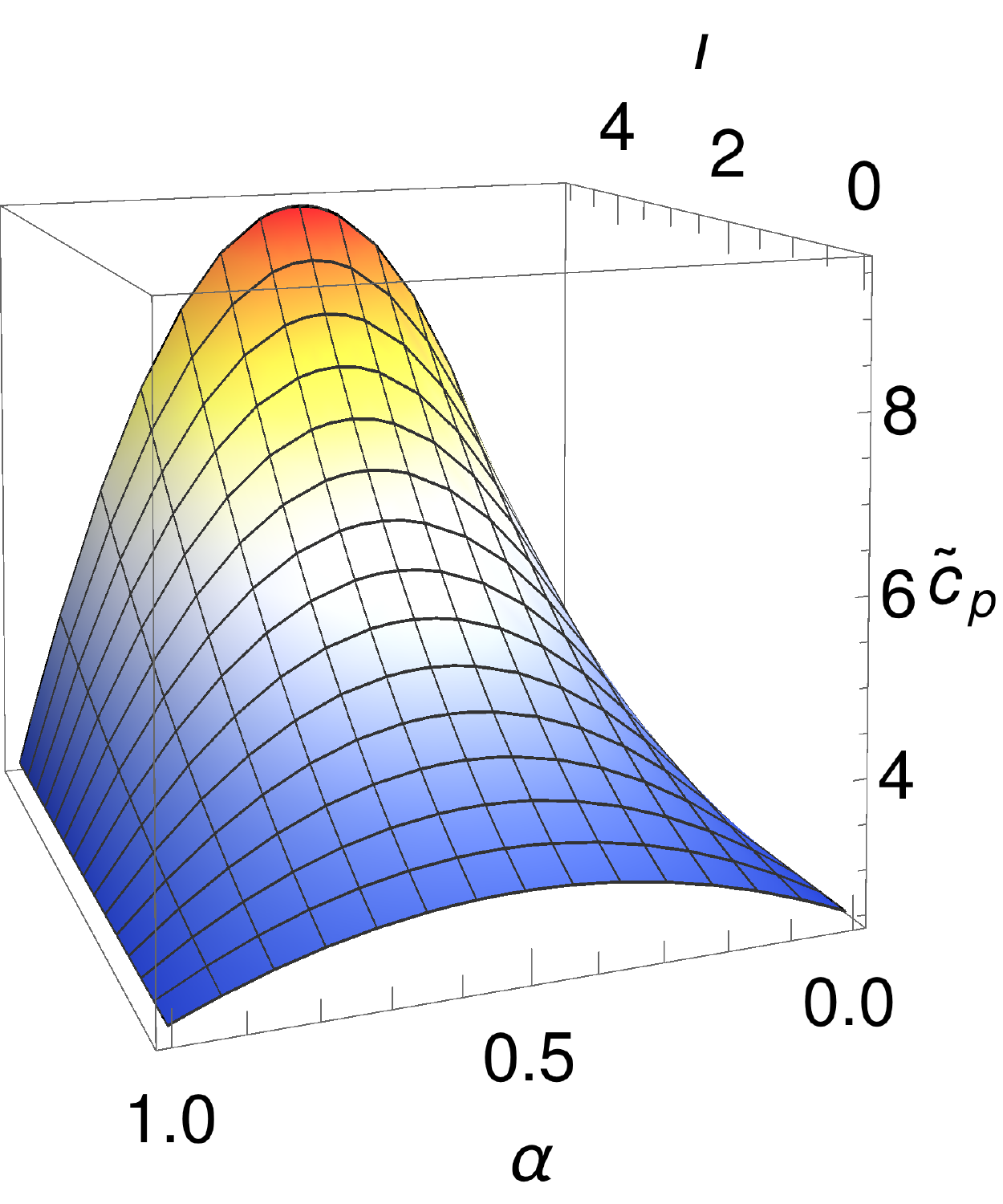}
\caption{Heat capacity per particle at constant pressure 
$\tilde{c}_p(\iota,\alpha)$ as a
function of $\alpha$ and $\iota$.
One can see significant increase of the heat capacity at small temperatures related to energy of ionization of the plasma.
}
\label{Fig:c_p}
\end{figure}

\begin{figure}[h]
\includegraphics[scale=0.3]{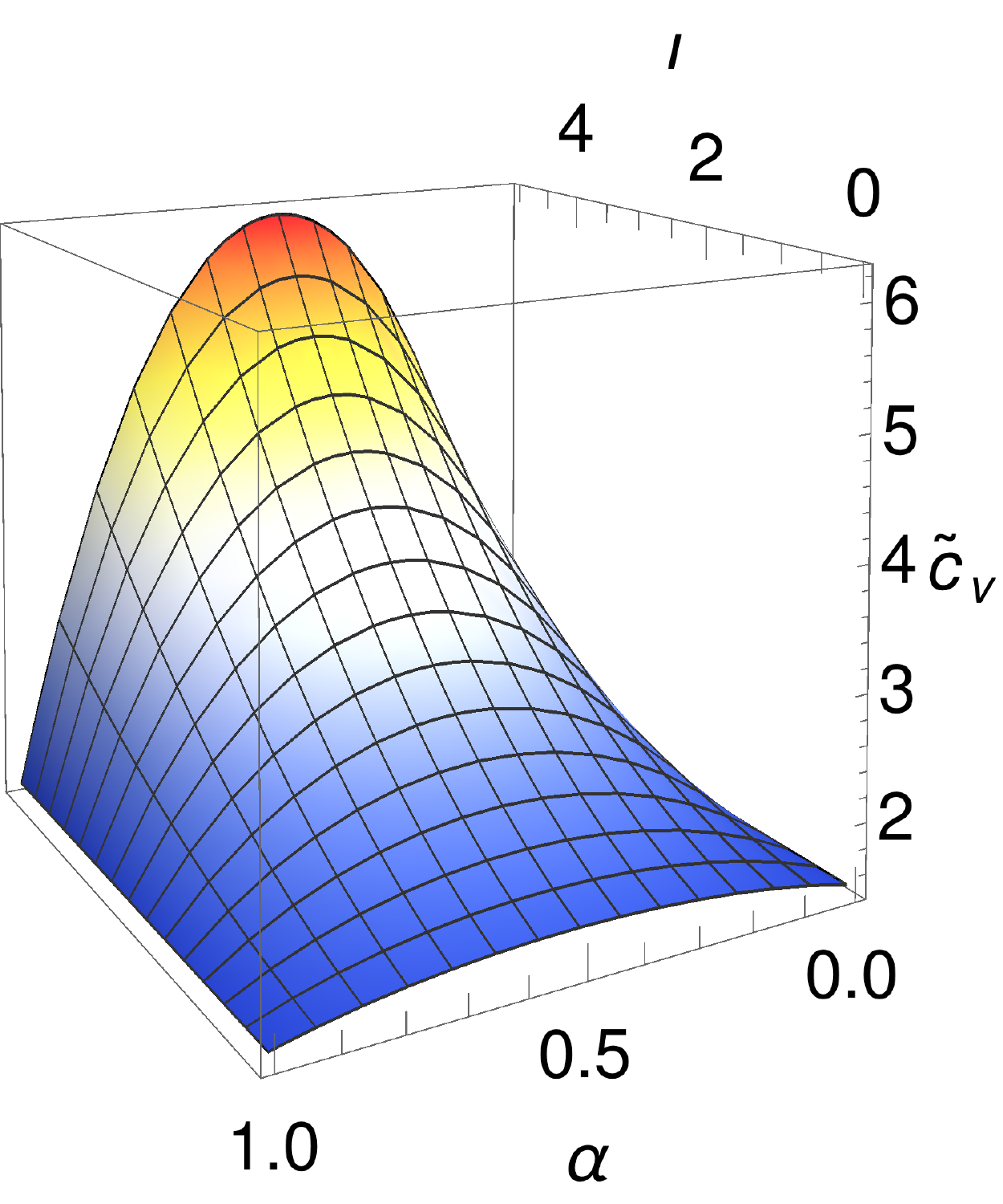}
\caption{Heat capacity per particle of partially ionized plasma at constant volume $\tilde{c}_v(\iota,\alpha)$}.
\label{Fig:c_v}
\end{figure}

Very often for low enough densities the correlation energy is negligible
and the pressure $p=n_\mathrm{tot}T$ is determined by the total volume density 
$n_\mathrm{tot}=n_0+n_p+n_e$ of all particles: neutral hydrogen atoms $n_0$,
protons $n_p$ and electrons $n_e$.
As the electron mass $m$ is much smaller than the proton mass $M$
for the mass density we have $\rho=M n_\rho$, where $n_\rho=n_0+n_p$.
The equality of the chemical potential of the particles $\mu_0=\mu_p+\mu_e$
for the reaction H~$\leftrightarrow$~e + p
gives the Saha equation which applied for the hydrogen plasma gives for the degree of ionization
\begin{align}&\nn
\alpha\equiv \frac{n_p}{n_\rho}=\frac1{\sqrt{1+p/p_\mathrm{_S}}},\quad
\frac{p}{p_\mathrm{_S}}=\frac1{\alpha^2}-1,\quad 
\frac{n_\rho}{n_\mathrm{_S}}=\frac{1-\alpha}{\alpha^2},
\\&
p_\mathrm{_S}\equiv n_\mathrm{_S}T,\quad
n_\mathrm{_S}\equiv n_q\mathrm{e}^{-\iota},\quad\iota\equiv\frac{I}{T},\quad
n_q=\left(\frac{mT}{2\pi\hbar^2}\right)^{\!\! 3/2},\nn
\end{align}
where $I$ is the hydrogen ionization potential and $\iota$ is reciprocal temperature in ionization potential units.
For the partially ionized hydrogen plasma the internal energy is given by the sum
of the kinetic energy of the particles and the energy of the ionization
\begin{align}&
\varepsilon=\left(c_v Tn_\mathrm{tot}+I n_e\right)/\rho,\quad w=\varepsilon+p/\rho,
\quad n_\rho=\frac{\rho}{M},
\\&\nonumber
n_e=n_p=\alpha n_\rho, \quad n_0=(1-\alpha)n_\rho,\quad
n_\mathrm{tot}=(1+\alpha)n_\rho,\\&\nonumber
c_v\equiv 3/2, \quad c_ p\equiv c_v+1=5/3,\quad \gamma_a\equiv c_p/c_v=5/3.
\end{align}
Using
\begin{align}&
T\left(\frac{\partial\alpha}{\partial T}\right)_{\!\!p}
=\frac{(1-\alpha^2)\alpha}{2}\,(c_p+\iota),\\&
T\left(\frac{\partial\alpha}{\partial T}\right)_{\!\!\rho}
=\frac{(1-\alpha)\alpha}{2-\alpha}\,(c_v+\iota)
\end{align}
the differentiation of the thermodynamic potentials after some algebra gives
\begin{align}&
\tilde{c}_p\equiv
\frac{\rho\,\mathcal{C}_p}{n_\mathrm{tot}}
=c_p+\frac12 (c_p+\iota)^2\,(1-\alpha)\alpha,\\&
\tilde{c}_v\equiv\frac{\rho\,\mathcal{C}_v}{n_\mathrm{tot}}
=c_v+\frac{(c_v+\iota)^2\, (1-\alpha)\alpha}
{2+(1-\alpha)\alpha},\\&
\gamma=\dfrac{\mathcal{C}_p}{\mathcal{C}_v}
=\frac{c_p+\dfrac12 (c_p+\iota)^2\,(1-\alpha)\alpha}
{c_v+\dfrac{(c_v+\iota)^2\,\alpha(1-\alpha)}
{2+(1-\alpha)\alpha}}.
\label{gamma}
\end{align}
The corresponding dimensionless functions from the dimensionless arguments 
$\alpha$ and $\iota$ are depicted in 
Figs.~\ref{Fig:gamma},\ref{Fig:c_p} and \ref{Fig:c_v}.
Perhaps it is specific only for hydrogen plasma that the functions
depend on the ionization degree only by the combination 
$\varphi\equiv (1-\alpha)\alpha/2$ and 
are symmetric with respect of the exchange 
$\alpha\leftrightarrow(1-\alpha)$.
The coefficient 
$\rho/n_\mathrm{tot}=\left<M\right>=M/(1+\alpha)$
is just the averaged mass of the cocktail.
We consider that the model consideration of the hydrogen plasma
reveals common features for all plasmas: for temperatures smaller than ionization potential the heat capacities 
$\tilde{c}_p$ and $\tilde{c}_v$ are created mainly by the ionization processes and
for $\alpha \approx 1/2$ they can significantly exceed  atomic values $c_p$ and $c_v$.
This leads that the polytropic index can differ from its atomic value
and relative value $\gamma/\gamma_a$ can be bigger or smaller than one as we can see from Fig.~\ref{Fig:gamma},

At fixed mass density $\rho$ the return to usual variables is simple: $T=I/\iota$
and $p=(1+\alpha)T\rho/M=T\rho/\left<M\right>$.
These results for hydrogen plasma reveal 
that even for significant degree of ionization
$\alpha\simeq 1$ 
when temperatures are comparable with the ionization potential,
the polytropic index $\gamma$ can have significant deviation
from single atomic value $\gamma_a$ which is exactly reached 
for $(1-\alpha)\alpha=0$.

In conclusion, we suggest that for interpretation of the experimental data for the adiabatic processes $\left(\partial T/\partial\rho\right)_s$ is indispensable to take into account 
ionization recombination processes even if the correlation energy is negligible.
For the pure argon plasma all energy levels are well-known and it 
is worthwhile to calculate theoretical curve passing near open circles 
in Fig.~5 of the commented article\cite{Takahashi:18} revealing 
$n_e-T_\mathrm{eff}$ correlation and effective polytropic index
$\gamma$. 
For and ideal gas cocktail adiabatic expansion means
conservation of the entropy per unit mass when Saha 
equation is taken into account.
It is a doable task and we consider as extremely interesting
experimental data presented Fig.~5 to be accompanied
with a state of the art theoretical curve
which describes deviation from simple polytropic model.

\textit{Apropos:}
The experimental set-up presented in Fig.~1 of the commented article\cite{Takahashi:18} remains a propulsion engine of a magneto-plasma rocket.
We use the opportunity to mention 
a new idea that not only helicon waves 
but antennas exciting Alfv\'en waves (AW) 
can be even the better solution
for heating of hot dense plasma by viscosity friction. 
The area of of AW damping will be similar to the combustion chamber of chemical jet engines. 
And creation of propulsion will be analogous to the launching of solar wind by absorption of AW as Hannes Alfv\`en suggested many years ago.

\end{document}